\documentstyle[preprint,aps]{revtex}

\makeatletter

\def\compoundrel#1\over#2{\mathpalette\compoundreL{{#1}\over{#2}}}
\def\compoundreL#1#2{\compoundREL#1#2}
\def\compoundREL#1#2\over#3{\mathrel
  {\vcenter{\hbox{$\m@th\buildrel{#1#2}\over{#1#3}$}}}}
\makeatother
\newcommand{\be}{\begin{equation}}
\newcommand{\ee}{\end{equation}}
\newcommand{\bea}{\begin{eqnarray}}
\newcommand{\eea}{\end{eqnarray}}
\newcommand{\bref}[1]{(\ref{#1})}
\newcommand{\mapright}[1]{%
	\smash{\mathop{%
	\hbox to 1cm{\rightarrowfill}}\limits^{#1}}}

\begin{document}
\draft
\title{{Assignments of Universal Texture Components \\ for Quark and Lepton  Mass Matrices}}

\author{Koichi MATSUDA}
\address{
Graduate school of Science, 
Osaka University, Toyonaka, Osaka 560-0043, Japan}
\author{Hiroyuki NISHIURA}
\address{
Department of General Education, 
Junior College of Osaka Institute of Technology, \\
Asahi-ku, Osaka 535-8585, Japan}

\date{August 30, 2003}
\maketitle

\begin{abstract}

We reanalyze the mass matrices model of quarks and leptons which gives a unified description of 
quark and lepton mass matrices with the same texture form. 
By investigating possible types of the assignment for the texture's 
components of this mass matrix form, we find that a different assignment 
for up-quarks from one for down-quarks can lead to consistent values of 
CKM mixing matrix. 
This finding overcomes a weak point of the previous analysis of the model. 
We also obtain some relations among the CKM mixing matrix parameters, 
which are independent of evolution effects.
\end{abstract}
\pacs{PACS number(s): 12.15.Ff, 11.30.Hv}


Recent neutrino oscillation experiments\cite{skamioka} have highly
suggested a nearly bimaximal lepton mixing $(\sin^2 2\theta_{12}\sim 1$, 
$\sin^2 2\theta_{23}\simeq 1)$ in contrast with the small quark mixing.
In order to reproduce these lepton and quark mixings, 
mass matrices of various structures with texture zeros has been investigated in the literature. 
Symmetric or Hermitian six and five texture zero models were systematically discussed 
by Ramond et al.\cite{Ramond}. 
Before the work of Ramond et al., Fritzsch\cite{fritzsch}  proposed a six texture zero model, 
nonsymmetric or non-Hermitian six texture zero quark mass matrices model[nearest-neighber-interaction(NNI) model] 
was proposed by Branco, Lavoura, and Mota\cite{branco}. 
Subsequently, Hermitian four texture model dealing with the quark and lepton mass matrices 
on the same footing are discussed by many authors\cite{du}. 
Phenomenological quark mass matrices have been discussed from various points of view\cite{shizuoka}.   
Recently a mass matrix model has been proposed\cite{Koide} 
with the following universal structure for all quarks and leptons 
\begin{equation}
\left(
\begin{array}{lll}
\ 0 & \ A & \ A \\
\ A & \ B & \ C \\
\ A & \ C & \ B \\
\end{array}
\right). \ \ 
\end{equation}
This form has been originally used for leptons(neutrinos) 
in order to reproduce a nearly bimaximal lepton mixing\cite{Fukuyama}--\cite{Nishiura}. 
Nevertheless it turns out that this model also leads to reasonable values of CKM quark mixing\cite{Koide}.   
Unfortunately, however, the model in Ref\cite{Koide} leads to $\frac{|V_{ub}|}{|V_{cb}|}=\sqrt{\frac{m_u}{m_c}}$, 
and consequently to some what small predicted value for $|V_{ub}|=0.0020 - 0.0029$  
with respect to the present experimental value $|V_{ub}|=0.0036  \pm 0.0007$\cite{PDG}.
This is caused by the assignment used in ref\cite{Koide} for texture's components(A,B, and C) of the mass matrix  
in which $A=\pm\sqrt{\frac{m_2m_1}{2}}$, $B=\frac{1}{2} \left(m_3+m_2-m_1\right)$ and 
$C=-\frac{1}{2}\left(m_3-m_2+m_1\right)$ with the i-th genaration mass $m_i$ are proposed 
both for up- and down- quarks.
In the present paper, we concentrate ourselves on this type of mass matrix model and explore every new possible assignments for texture's components(A, B, and C). 
We will point out that there exists another possible new assignment for A, B, and C  
and that a different type of the assignment for up-quarks from the one for down-quarks can lead to 
consistent values of CKM mixing matrix. 
Namely we show the combination of the new assignment that 
$A=\pm\sqrt{\frac{m_3m_1}{2}}$, $B=\frac{1}{2} \left(m_3+m_2-m_1\right)$ and 
$C=\frac{1}{2}\left(m_3-m_2-m_1\right)$ for up-quarks, 
and previously proposed one that $A=\pm\sqrt{\frac{m_2m_1}{2}}$, $B=\frac{1}{2} \left(m_3+m_2-m_1\right)$ and 
$C=-\frac{1}{2}\left(m_3-m_2+m_1\right)$ for down- quarks is favorable 
to reproduce the consistent values of CKM quark mixing. 
This new finding of the another assignment overcomes the above weak 
point of the approach in Ref\cite{Koide}. 
The new and positive feature of this work is presented in Table 1.
\par
Our mass matrices 
\(M_u\), \(M_d\), \(M_\nu\) and  \(M_e\) 
(mass matrices of up quarks (\(u,c,t\)), down quarks (\(d,s,b\)), 
neutrinos (\(\nu_e,\nu_\mu,\nu_\tau\)) and 
charged leptons (\(e,\mu,\tau\)), 
respectively) are given as follows:\cite{Koide}
\begin{equation}
M_f = P_{f}^\dagger\widehat{M}_fP_{f}^\dagger, \
\quad \quad \quad \quad 
\end{equation}
with
\begin{equation}
\widehat{M}_f=
\left(
\begin{array}{lll}
\ 0 & \ A_f & \ A_f \\
\ A_f & \ B_f & \ C_f \\
\ A_f & \ C_f & \ B_f \\
\end{array}
\right) \ \, \left(f=u,d,\nu,e\right),
\label{texture}
\end{equation}
where $P_{Lf}$ is the diagonal phase matrices 
and $A_f$, $B_f$, and $C_f$ are real parameters. 
This structure of mass matrix was previously suggested and used 
for the neutrino mass matrix in Refs\cite{Fukuyama}--\cite{Nishiura},  
using the basis where the charged-lepton mass matrix is diagonal, 
motivated by the experimental finding of maximal \(\nu_\mu\)--\(\nu_\tau\) 
mixing\cite{skamioka}.
\par  
Hereafter, for brevity, we will omit the flavour index.
The eigen-masses of Eq. \bref{texture} are given by
$
\frac{1}{2}\left(B+C-\sqrt{8A^2 + (B+C)^2}\right)$ ,
$\frac{1}{2}\left(B+C+\sqrt{8A^2 + (B+C)^2}\right)$ , and
$(B-C)$. 
Therefore, there are three types of assignments for the texture's 
components of \(\widehat{M}\) according to the assignments for the eigen-mass $m_i$ :
\par
(i)Type A: 
\begin{eqnarray}
-m_1& =&
\frac{1}{2}
\left(B+C-\sqrt{8A^2 + (B+C)^2}
\right) ,\\
m_2& =&\frac{1}{2}
\left(B+C+\sqrt{8A^2 + (B+C)^2}
\right) ,\\
m_3& =&B-C. 
\end{eqnarray}
This is the case that $B-C$ is the largest value. In this type, 
the texture's components of \(\widehat{M}\) are expressed 
in terms of $m_i$ as 
\begin{eqnarray}
A & =&\pm\sqrt{\frac{m_2m_1}{2}}  ,\nonumber\\
B & =&\frac{1}{2} \left(m_3+m_2-m_1\right) ,\label{eq2003}\\
C & =&-\frac{1}{2}\left(m_3-m_2+m_1\right) .\nonumber 
\end{eqnarray}
That is, 
\(\widehat{M}\) is diagonalized by an orthogonal matrix \(O\) as
\begin{equation}
O^T\widehat{M}O=
\left(
	\begin{array}{ccc}
	-m_1 & 0 & 0\\
	0 & m_2 & 0\\
	0 &  0 & m_3
	\end{array}
\right),
\end{equation}
with
\begin{equation}
O\equiv
\left(
\begin{array}{ccc}
{ \pm c}&
{ \pm s}&
{0} \\
{-\frac{s}{\sqrt{2}}}&
{\frac{c}{\sqrt{2}}}&
{-\frac{1}{\sqrt{2}}} \\
{-\frac{s}{\sqrt{2}}}&
{\frac{c}{\sqrt{2}}}&
{\frac{1}{\sqrt{2}}}
\end{array}
\right). \label{eq990114} \\
\label{O}
\end{equation}
Here \(c\) and \(s\) are defined by $c=\sqrt{\frac{m_2}{m_2+m_1}}$ and $s=\sqrt{\frac{m_1}{m_2+m_1}}$.
It should be noted that the elements of \(O\) are independent of \(m_3\) 
because of the above structure of \(\widehat{M}\).
This type A is used in Ref\cite{Koide}. 
\par
(ii)Type B: This assignment is obtained by exchanging $m_2$ and $m_3$ in Type A.
\begin{eqnarray}
-m_1& =&
\frac{1}{2}
\left(B+C-\sqrt{8A^2 + (B+C)^2}
\right) ,\\
m_2& =&B-C,\\
m_3& =&\frac{1}{2}
\left(B+C+\sqrt{8A^2 + (B+C)^2}
\right). 
\end{eqnarray}
In this type, the texture's components of \(\widehat{M}\) are expressed as 
\begin{eqnarray}
A & =&\pm\sqrt{\frac{m_3m_1}{2}}  ,\nonumber\\
B & =&\frac{1}{2}\left(m_3 +m_2-m_1\right) ,\label{eq20032}\\
C & =&\frac{1}{2}\left(m_3-m_2-m_1\right) .\nonumber 
\end{eqnarray}
The orthogonal matrix \(O^\prime\) which diagonalizes 
\(\widehat{M}\) ($(O^{\prime T}\widehat{M}O^\prime=diag(-m_1,m_2,m_3)$)is 
given by 
\begin{equation}
O^\prime\equiv
\left(
\begin{array}{ccc}
{\pm c^\prime}&
{0}&
{\pm s^\prime} \\
{-\frac{s^\prime}{\sqrt{2}}}&
{-\frac{1}{\sqrt{2}}}&
{\frac{c^\prime}{\sqrt{2}}} \\
{-\frac{s^\prime}{\sqrt{2}}}&
{\frac{1}{\sqrt{2}}}&
{\frac{c^\prime}{\sqrt{2}}}
\end{array}
\right). \label{eq990114B} \\
\label{OB}
\end{equation}
Here \(c^\prime\) and \(s^\prime\) are defined by $c^\prime=\sqrt{\frac{m_3}{m_3+m_1}}$ and $s^\prime=\sqrt{\frac{m_1}{m_3+m_1}}$.
\par
(iii)Type C: This assignment is obtained by exchanging $m_1$ for $m_2$ in Type B. 
However, this type is not useful in the discussion on the quark sector.
\par
Taking the type A assignment of mass matrices both for up and down quarks,
the authors in Ref \cite{Koide} have discussed the quark mixing matrix 
and obtained the following prediction which is almost independent of the 
RGE effects,
\begin{equation}
\frac{|V_{ub}|}{|V_{cb}|}=\sqrt{\frac{m_u}{m_c}} = 0.051 - 0.067.
\label{mat092201}
\end{equation}
By substituting the experimental values 
\(|V_{cb}|_{exp}\)\(=\)\(0.0412 \pm 0.020\) \cite{PDG} 
into Eq.(\ref{mat092201}), one obtain
\begin{equation}
|V_{ub}|= \sqrt{\frac{m_u}{m_c}} |V_{cb}|_{exp} = 0.0020 - 0.0029.
\end{equation}
However, this value is somewhat smaller than
the present experimental value $|V_{ub}|=0.0036  \pm 0.0007$\cite{PDG}. 
\par
In this paper, taking the type B assignment for up quarks and the type A 
for down quarks, we reanalyze the quark mixing matrix of the model. 
In this assignment, \(M_u\) and \(M_d\) have the same zero texture with different assignments as follows:
\begin{eqnarray}
&&M_{u}=
P_u^\dagger\left(
	\begin{array}{ccc}
	 0   \quad & \pm\sqrt{\frac{m^0_tm^0_u}{2}}\quad & \pm\sqrt{\frac{m^0_tm^0_u}{2}} \\
	\pm\sqrt{\frac{m^0_tm^0_u}{2}} 
        \quad& \frac{1}{2}\left(m^0_t +m^0_c-m^0_u\right) 
        \quad& \frac{1}{2}\left(m^0_t-m^0_c-m^0_u\right)\\
	 \pm\sqrt{\frac{m^0_tm^0_u}{2}}   
        \quad & \frac{1}{2}\left(m^0_t-m^0_c-m^0_u\right)
        \quad & \frac{1}{2}\left(m^0_t +m^0_c-m^0_u\right)
	\end{array}
\right) P_u^\dagger, 
\nonumber\\
&&M_{d}=
P_d ^\dagger\left(
	\begin{array}{ccc}
	 0  \quad& \pm\sqrt{\frac{m^0_sm^0_d}{2}}
         \quad &  \pm\sqrt{\frac{m^0_sm^0_d}{2}} \\
	\pm\sqrt{\frac{m^0_sm^0_d}{2}}
        \quad & \frac{1}{2} \left(m^0_b+m^0_s-m^0_u\right)
        \quad& -\frac{1}{2}\left(m^0_b-m^0_s+m^0_d\right)\\
	 \pm\sqrt{\frac{m^0_sm^0_d}{2}} 
        \quad& -\frac{1}{2}\left(m^0_b-m^0_s+m^0_d\right)
        \quad& \frac{1}{2} \left(m^0_b+m^0_s-m^0_u\right)
	\end{array}
\right) P_d^\dagger \label{eq5}.
\end{eqnarray}
where \(P_u\) and \(P_d\) are the $CP$ violating phase factors.
These quark mass matrices $M_f = P_{f}^\dagger\widehat{M}_fP_{f}^\dagger  \ \ (f=u,d)$ 
are diagonalized by the bi-unitary transformation
\begin{equation}
D_f  =  U_{Lf}^\dagger M_f U_{Rf}\ , 
\end{equation}
where $U_{Lu}\equiv P_u^\dagger O_u^\prime$, $U_{Ru}\equiv P_u O_u^\prime$, 
$U_{Ld}\equiv P_d^\dagger O_d$, and $U_{Rd}\equiv P_d O_d$. 
Here $O_u^\prime$ and $O_d$ are given by Eq.~(\ref{OB}) and Eq.~(\ref{O}), respectively. 
Then, the Cabibbo--Kobayashi--Maskawa (CKM) \cite{CKM} quark mixing 
matrix \(V\) is given by
\begin{eqnarray}
V&=&U^\dagger_{Lu}U_{Ld}=O^{\prime T}_uP_{u}P^\dagger_{d} O_d\nonumber\\[.1in]
& =&
\left(
\begin{array}{ccc}
c^\prime_uc_d+\rho s^\prime_u s_d \quad & c^\prime_u s_d-\rho s^\prime_u c_d 
\quad & -{\sigma}s^\prime_u \\
-{\sigma}s_d \quad & {\sigma}c_d \quad & \rho \\
s^\prime_u c_d-{\rho}c^\prime_u s_d \quad & s^\prime_u s_d+{\rho}c^\prime_u c_d 
\quad & {\sigma}c^\prime_u \\
\end{array}
\right),\label{eq-ourckm} 
\end{eqnarray}
where  \(\rho\) and \(\sigma\) are defined by 
\begin{equation}
\rho=\frac{1}{2}(e^{i\delta_3}+e^{i\delta_2})
=\cos\frac{\delta_3 - \delta_2}{2} \exp i
\left( \frac{\delta_3 + \delta_2}{2} \right) \ ,
\end{equation}
\begin{equation}
\sigma=\frac{1}{2}(e^{i\delta_3}-e^{i\delta_2})
= \sin\frac{\delta_3 - \delta_2}{2} 
\exp i \left( \frac{\delta_3 + \delta_2}{2}+ \frac{\pi}{2}
\right) \ . 
\end{equation}
Here we have put \(P \equiv P_{u}P^\dagger_{d} \equiv 
\mbox{diag}(e^{i\delta_1}, e^{i\delta_2},e^{i\delta_3})\), and
we have taken \(\delta_1=0\) without 
loss of generality.

Then, the explicit magnitudes of the components of \(V\) are expressed as
\begin{eqnarray}
\left|V^0_{cb}\right| & =&\left|\rho\right| 
= \cos\frac{\delta_3-\delta_2}{2},\label{eq3021}\\
\left|V^0_{ub}\right|& =&\left|\sigma\right| s^\prime_u 
= \frac{\sin\frac{\delta_3-\delta_2}{2}}{\sqrt{1+m^0_u/m^0_t}}\sqrt{
\frac{m^0_u}{m^0_t}},\label{eq3022}\\
\left|V^0_{cd}\right|& =&\left|\sigma\right| s_d 
= \frac{\sin\frac{\delta_3-\delta_2}{2}}{\sqrt{1+m^0_d/m^0_s}}\sqrt{
\frac{m^0_d}{m^0_s}},\label{eq3024}\\
\left| V^0_{us}\right|
& =&c^\prime_u s_d\left|1 \mp \rho{\frac{s^\prime_u}{c^\prime_u}}{\frac{c_d}{s_d}}\right|
=\sqrt{\frac{m^0_t}{m^0_t+m^0_u}} \sqrt{\frac{m^0_d}{m^0_s+m^0_d}}\nonumber \\
& \times&
\left[1 \mp 2\cos\frac{\delta_3-\delta_2}{2}
\cos\frac{\delta_3+\delta_2}{2}
 \sqrt{\frac{m^0_u m^0_s}{m^0_t m^0_d}}
+ \cos^2 \frac{\delta_3 - \delta_2}{2}
\left( \frac{m^0_u m^0_s}{m^0_t m^0_d}\right)\right]^{\frac{1}{2}},\label{eq3025}\\
\left| V^0_{td}\right|
& =&c^\prime_u s_d\left|\rho \mp {\frac{s^\prime_u}{c_u}}{\frac{c^\prime_u}{s_d}}\right|
=\sqrt{\frac{m^0_t}{m^0_t+m^0_u}} \sqrt{\frac{m^0_d}{m^0_s+m^0_d}}\nonumber \\
& \times&
\left[\cos^2 \frac{\delta_3 - \delta_2}{2} \mp 2\cos\frac{\delta_3-\delta_2}{2}
\cos\frac{\delta_3+\delta_2}{2} \sqrt{\frac{m^0_u m^0_s}{m^0_t m^0_d}}
+ 
\left( \frac{m^0_u m^0_s}{m^0_t m^0_d}\right) 
\right]^{\frac{1}{2}},\label{eq3026}\\
\left|V^0_{ts}\right|
& =&c^\prime_u c_d\left|\rho \pm {\frac{s^\prime_u}{s_u}}{\frac{c^\prime_u}{c_d}}\right|
=\sqrt{\frac{m^0_t}{m^0_t+m^0_u}} \sqrt{\frac{m^0_s}{m^0_s+m^0_d}}\nonumber \\
& \times&
\left[\cos^2 \frac{\delta_3 - \delta_2}{2} \pm2\cos\frac{\delta_3-\delta_2}{2}
\cos\frac{\delta_3+\delta_2}{2} \sqrt{\frac{m^0_u m^0_d}{m^0_t m^0_s}}
+ 
\left( \frac{m^0_u m^0_d}{m^0_t m^0_s}\right) 
\right]^{\frac{1}{2}}.\label{eq3023}
\end{eqnarray}
It should be noted that $|V^0_{us}|$, $|V^0_{td}|$, and $|V^0_{ts}|$ are 
almost independent of $(\delta_3 + \delta_2)$ and they are given from 
Eq.~(\ref{eq3025})-(\ref{eq3023}) as
\begin{eqnarray}
\left| V^0_{us}\right|
& \simeq&\sqrt{\frac{m^0_d}{m^0_s}},\label{eq30257}\\
\left| V^0_{td}\right|
& \simeq&\sqrt{\frac{m^0_d}{m^0_s}}\cos\frac{\delta_3 - \delta_2}{2}, \label{eq30267}\\
\left|V^0_{ts}\right|
& \simeq&\cos\frac{\delta_3 - \delta_2}{2}.\label{eq30237}
\end{eqnarray}
Therefore, the independent parameters in the expression $|V^0_{ij}|$ are 
$\theta^\prime_u = \tan^{-1}(m^0_u/m^0_t)$,
$\theta_d = \tan^{-1} (m^0_d/m^0_s)$, and ($\delta_3-\delta_2$). 
Among them, the two parameters $\theta^\prime_u$ and $\theta_d$ are already 
fixed by the quark masses. 
Therefore, the present model has an adjustable parameter $(\delta_3 -\delta_2)$ 
in $|V^0_{ij}|$, which is fixed to reproduce the observed CKM matrix 
parameters at $\mu=m_z$\cite{PDG}:
\begin{eqnarray}
&&|V_{us}|_{\mbox{{\tiny exp}}}= 0.2196 \pm 0.0026, \quad 
|V_{cb}|_{\mbox{{\tiny exp}}}= 0.0412 \pm 0.0020, \nonumber\\
&&|V_{ub}|_{\mbox{{\tiny exp}}}= (3.6 \pm 0.7)\times 10^{-3}, \quad
\label{mat091902}
\end{eqnarray}
\par
The relations in Eqs.~(\ref{eq3021})--(\ref{eq3023}) hold only at the unification scale $\mu=M_X$. 
So we now consider evolution effects. As is well known \cite{evol}, 
the evolution effects are approximately described as
\begin{eqnarray}
\frac{m^0_d/m^0_b}{m_d/m_b}
 & \simeq&\frac{m^0_s/m^0_b}{m_s/m_b} 
   \simeq \frac{\left|V^0_{ub}\right|}{\left|V_{ub}\right|} 
   \simeq \frac{\left|V^0_{cb}\right|}{\left|V_{cb}\right|} 
   \simeq \frac{\left|V^0_{td}\right|}{\left|V_{td}\right|} 
   \simeq \frac{\left|V^0_{ts}\right|}{\left|V_{ts}\right|} 
   \simeq 1+\epsilon_d,\\
\frac{m^0_u/m^0_c}{m_u/m_c}
 & \simeq&\frac{m^0_d/m^0_s}{m_d/m_s} 
   \simeq \frac{\left|V^0_{us}\right|}{\left|V_{us}\right|} 
   \simeq \frac{\left|V^0_{cd}\right|}{\left|V_{cd}\right|} 
   \simeq 1.
\end{eqnarray}
where $m^0_q$ and $V^0_{ij}$ ($m_q$ and $V_{ij}$) denote the values 
at $\mu=M_X$($\mu=m_Z$). In the following numerical calculations, 
we use the running quark mass at \(\mu=m_Z\)  and  at \(\mu=M_X\) \cite{Fusaoka}:
\begin{equation}
\begin{array}{lll}
m_u(m_Z)=2.33^{+0.42}_{-0.45}\ \mbox{MeV},& 
m_c(m_Z)=677^{+56}_{-61}\ \mbox{MeV},& 
m_t(m_Z)=181 \pm 13\ \mbox{GeV},\\
m_d(m_Z)=4.69^{+0.60}_{-0.66}\ \mbox{MeV},& 
m_s(m_Z)=93.4^{+11.8}_{-13.0}\ \mbox{MeV},& 
m_b(m_Z)=3.00 \pm 0.11\ \mbox{GeV}.
\end{array}
\label{eq123103}
\end{equation}
\begin{equation}
\begin{array}{lll}
m_u(M_X)=1.04^{+0.19}_{-0.20}\ \mbox{MeV},& 
m_c(M_X)=302^{+25}_{-27}\ \mbox{MeV},& 
m_t(M_X)=129^{+196}_{-40}\ \mbox{GeV},\\
m_d(M_X)=1.33^{+0.17}_{-0.19}\ \mbox{MeV},& 
m_s(M_X)=26.5^{+3.3}_{-3.7}\ \mbox{MeV},& 
m_b(M_X)=1.00 \pm 0.04\ \mbox{GeV}.
\end{array}
\label{eq1231032}
\end{equation}
\par
First we note that the predictions
\begin{eqnarray}
\left|V^0_{us}\right|& \simeq&c^\prime_u s_d
\simeq \sqrt{\frac{m^0_d}{m^0_s}} \simeq 0.224, \label{eq30268}\\
\frac{\left|V^0_{td}\right|}{\left|V^0_{ts}\right|}& \simeq&\frac{s_d}{c_d}
=\sqrt{\frac{m^0_d}{m^0_s}} \simeq 0.224, \ \label{eq30269}
\end{eqnarray}
are almost independent of the RGE effects, because they do not
contain the phase difference, $(\delta_3 -\delta_2)$, which
is highly dependent on the energy scale as we discussed in the previous analysis 
and we know that the ratio $m_d/m_s$ is almost independent of the RGE effects.
We also obtain the relation
\begin{equation}
\frac{\left|V^0_{ub}\right|}{\left|V^0_{cd}\right|}
=\sqrt{\frac{(m^0_s+m^0_d)m^0_u}{(m^0_t+m^0_u)m^0_d}},
\end{equation}
which is independent of the phase difference.
\par

Next let us fix the parameters $\delta_3-\delta_2$ using the
expression of Eq.~(\ref{eq3021}) which  holds at $\mu=M_X$.
From Eq.~(\ref{eq3021}), with taking $\epsilon_d=\frac{m^0_d/m^0_b}{m_d/m_b}-1=-0.149$, we have
\begin{equation}
\cos\frac{\delta_3-\delta_2}{2} = \left|V_{cb}
\right|_{\mbox{{\tiny exp}}}(1+\epsilon_d)=(0.0412 \pm 0.0020 )(1+\epsilon_d)\ ,\end{equation}
\begin{equation}
\delta_3 -\delta_2 = 175.98^\circ. \label{eq3032}
\end{equation}

Thus we obtain $V_{ij}$ at \(\mu=m_Z\) as follows
\begin{eqnarray}
|V_{ub}|& \simeq&|V^0_{ub}|\frac{1}{1+\epsilon_d}
=\sqrt{\frac{\frac{m^0_u}{m^0_t}}{1+\frac{m^0_u}{m^0_t}}}
 \sqrt{1-\left|V_{cb}\right|^2_{\mbox{{\tiny exp}}}(1+\epsilon_d)^2}
 \frac{1}{1+\epsilon_d}
\simeq0.0033,\label{eq3033}\\
|V_{cd}|& \simeq&|V^0_{cd}|
=\sqrt{\frac{\frac{m^0_d}{m^0_s}}{1+\frac{m^0_d}{m^0_s}}}
 \sqrt{1-\left|V_{cb}\right|^2_{\mbox{{\tiny exp}}}(1+\epsilon_d)^2}
\simeq 0.22 ,\label{eq30332}\\
|V_{ts}|& \simeq& |V^0_{ts}|\frac{1}{1+\epsilon_d} 
\simeq \left|V_{cb}\right|_{\mbox{{\tiny exp}}}
\simeq 0.041,\label{eq3034}\\
|V_{td}|& \simeq&|V^0_{td}|\frac{1}{1+\epsilon_d}
\simeq\sqrt{\frac{m^0_d}{m^0_s}}\left|V_{cb}\right|_{\mbox{{\tiny exp}}}
\simeq0.0092,\label{eq3035}
\end{eqnarray}
which are consistent with the present experimental data.
Therefore, the value of \((\delta_3-\delta_2)\) in Eq.~(\ref{eq3032}) 
is acceptable.
\par
The remaining parameter $(\delta_3+\delta_2)$ in this model remains 
a free parameter to be fixed by the observed CP-violating phase 
\(\delta\) in the standard representation of the CKM quark mixing matrix  
\begin{eqnarray}
V_{\rm std} &=& \mbox{diag}(e^{\alpha_1^u},e^{\alpha_2^u},e^{\alpha_2^u})  \ V \ 
\mbox{diag}(e^{\alpha_1^d},e^{\alpha_2^d},e^{\alpha_2^d}) \nonumber \\
&=&
\left(
\begin{array}{ccc}
c_{13}c_{12} & c_{13}s_{12} & s_{13}e^{-i\delta} \\
-c_{23}s_{12}-s_{23}c_{12}s_{13} e^{i\delta}
&c_{23}c_{12}-s_{23}s_{12}s_{13} e^{i\delta} 
&s_{23}c_{13} \\
s_{23}s_{12}-c_{23}c_{12}s_{13} e^{i\delta}
 & -s_{23}c_{12}-c_{23}s_{12}s_{13} e^{i\delta} 
& c_{23}c_{13} \\
\end{array}
\right) \ .
\label{stdrep}
\end{eqnarray}
Here, \(\alpha_i^q\) comes from the rephasing in the quark fields 
to make the choice of phase convention.
The \(\delta\) in  Eq.~(\ref{stdrep}) is expressed, in the present model, by 
\begin{equation}
\delta =-\frac{\delta_3+\delta_2}{2}\ ,
\end{equation}
so that from the observed value  $\delta_{\mbox{{\tiny exp}}}= 59^\circ \pm 13^\circ$ 
we can fix $\delta_3+\delta_2= -118^\circ \pm 26^\circ $. 

\par
Note that the mass matrix does not keep its texture form
at \(\mu = m_Z\) 
because (1,2) and (1,3) components are not so small.
Therefore, we consider the RGE effect more precisely.
In our model, the  \(|V_{cb}|\) and the
CP violating phase \(\delta\) are controllable parameters 
with the unobservable parameters \(\delta_2\) and \(\delta_3\),
so we can adjust these parameters to the center values at \(\mu = m_Z\).
On the other hand, \(|V_{us}|\) and \(|V_{ub}|\) are 
restricted in our model by mass regions in Eq.(\ref{eq123103}).
Under these conditions, we estimate the numerical variation of 
the CKM matrix elements 
by using the two-loop RGE (MSSM (\(\tan \beta = 10\)) case) 
for the Yukawa coupling constants \cite{Fusaoka}. 
Then, we obtain the following numerical results for the evolution effects: 
\begin{eqnarray}
|V_{us}^0| &=&\sqrt{\frac{m_d^0}{m_d^0+m_s^0}}= 0.19-0.25 \hspace{0.8cm} 
  \quad \to \quad |V_{us}| = 0.19-0.25,\\
|V_{ub}^0| &=&\sqrt{\frac{m_u^0}{m_u^0+m_t^0}}= 0.0020-0.0034 \quad \to \quad |V_{ub}|= 0.0027-0.0038,\\
|V_{cb}^0| &=& 0.035 \hspace{4.5cm} \quad \to \quad |V_{cb}| = 0.041,\\
\delta^0 \ \  &=& 59^\circ \hspace{4.85cm} 
  \quad \to \quad \delta = 59^\circ.
\end{eqnarray}
These values are consistent with the approximations in Eqs.(\ref{eq30268})
-(\ref{eq3035}) and also the experimental data in Eq.(\ref{mat091902}).

\par
In conclusion, we have reanalyzed the quark mixing matrix using the mass matrix model of Ref\cite{Koide} with the universal texture form. 
We use different types of assignments for $A_f$, $B_f$, and $C_f$ 
in $\widehat{M}_f$ for (f=u and d). 
Namely, the type A for $\widehat{M}_d$, while the type B for $\widehat{M}_u$ 
are considered in this paper. This is in contrast with the previous analysis 
in Ref\cite{Koide} with use of the same type for both, 
which leads to some what small predicted value for $|V_{ub}|$ 
compared with the experimental value. 
It is shown that the present model predicts consistent values of 
CKM mixing matrix and the above weak point of the previous model is overcome. 
We also have relations $\left|V_{us}\right|\simeq \sqrt{\frac{m_d}{m_s}}$ and 
$\frac{\left|V_{td}\right|}{\left|V_{ts}\right|} \simeq \sqrt{\frac{m_d}{m_s}}$ 
which are almost independent of RGE effects.

\vspace{5mm}
We are grateful to Y. Koide and H. Fusaoka for the useful comments.
This work of K.M. was supported by the JSPS Research Fellowships 
for Young Scientists, No. 3700.



\begin{table}[htbp]
\begin{tabular}{c|c|c|c} 
down \(\backslash\) up & Type A & Type B & Type C \\ \hline
Type A &
\parbox{4.8cm}{\begin{center} \(\left|\frac{U_{ub}}{U_{cb}}\right|
                                 = \sqrt{\frac{m_u^0}{m_c^0}}\alt \mbox{exp}  \) \\
                              \(\left|\frac{U_{td}}{U_{ts}}\right|
                                 = \sqrt{\frac{m_d^0}{m_s^0}}\simeq \mbox{exp}  \) 
\end{center}} &
\parbox{4.8cm}{\begin{center} \(|U_{us}| \simeq \sqrt{\frac{m_d}{m_d+m_s}}\simeq \mbox{exp} \) \\
                              \(\left|\frac{U_{cd}}{U_{cs}}\right|
                                 = \sqrt{\frac{m_d^0}{m_s^0}}\simeq \mbox{exp}  \) 

\end{center}} &
\parbox{4.8cm}{\begin{center}\(\left|\frac{U_{ud}}{U_{us}}\right| = \sqrt{\frac{m_d^0}{m_s^0}}
                                                                \gg \mbox{exp} \) 
\end{center}} \\ \hline
Type B &
\parbox{4.8cm}{\begin{center} \(|U_{us}| \le \sqrt{\frac{m_u^0}{m_u^0+m_c^0}}\ll \mbox{exp}  \) \\
                              \(\left|\frac{U_{us}}{U_{cs}}\right|
                                 = \sqrt{\frac{m_u^0}{m_c^0}} \ll \mbox{exp}  \)
\end{center}} &
\parbox{4.8cm}{\begin{center} \(|U_{us}| \le \sqrt{\frac{m_u^0}{m_u^0+m_t^0}}\ll \mbox{exp} \) \\
                              \(|U_{cd}| \le \sqrt{\frac{m_d^0}{m_d^0+m_b^0}}\ll \mbox{exp} \) 
\end{center}} &
\parbox{4.8cm}{\begin{center} \(|U_{ud}| \le \sqrt{\frac{m_d^0}{m_d^0+m_b^0}}\ll \mbox{exp} \) \\
                              \(|U_{cs}| \le \sqrt{\frac{m_c^0}{m_c^0+m_t^0}}\ll \mbox{exp} \) 
\end{center}} \\ \hline
Type C &
\parbox{4.8cm}{\begin{center} \(\left|\frac{U_{ud}}{U_{cd}}\right| = \sqrt{\frac{m_u^0}{m_c^0}}
                                                                \gg \mbox{exp} \) 
\end{center}} &
\parbox{4.8cm}{\begin{center} \(|U_{ud}| \le \sqrt{\frac{m_u^0}{m_u^0+m_t^0}}\ll \mbox{exp} \) \\
                            \(|U_{cs}| \le \sqrt{\frac{m_s^0}{m_s^0+m_b^0}}\ll \mbox{exp} \) 
\end{center}} &
\parbox{4.8cm}{\begin{center} \(|U_{cd}| \le \sqrt{\frac{m_c^0}{m_c^0+m_t^0}}\ll \mbox{exp} \) 
\end{center}} 
\end{tabular}

\caption{%
The results of quark mixing matrix element $U_{ij}$ are 
shown for three types of the assignment for up and down quarks mass matrices. 
These results are independent of the phases \(\delta_2\) and \(\delta_3\) 
and are useful for the consistency check 
between our model and the experiments
because these values are hardly changed by the RGE effects from the GUT scale to the EW scale.
The "exp" represents the corresponding experimental value of the left-hand side of inequality or equation.
Here Type A is proposed in Ref [7]. Type B and Type C are new assignments proposed in the present paper.  
We find that only a combination of Type B for up and Type A for down quarks leads to the well consistent 
CKM mixing matrix. 
Other combinations fail to reproduce consistent quark mixing 
because of the results indicated in this table. }
\label{table1}
\end{table}

\end{document}